# Modelling Purcell enhancement of metasurfaces supporting *quasi*-bound states in the continuum

Joshua T. Y. Tse[1,2]*, Taisuke Enomoto[1], Shunsuke Murai[2], and Katsuhisa Tanaka[1]

[1]Department of Material Chemistry, Graduate School of Engineering, Kyoto University, Katsura, Kyoto 615-8510, Japan

[2]Department of Physics and Electronics, Graduate School of Engineering, Osaka Metropolitan University, Osaka 599-8531, Japan

*Email: jtytse@omu.ac.jp

**ABSTRACT**

Bound states in the continuum (BIC) exhibit extremely high quality factors due to the lack of radiation loss and thus are widely studied for Purcell enhancement. However, a closer examination reveals that the enhancement is absent at the BIC due to the lack of out-coupling capability, but the strong enhancement is only observed at nearby configuration, namely *quasi*-BIC. To study this unique behavior of the Purcell enhancement near BIC, we built an analytical model with spectral parameters to analyze the Purcell enhancement on metasurfaces supporting *quasi*-BIC. Our analytical model predicts the average Purcell enhancement by metasurfaces coupled to a luminescent medium, utilizing parameters that are formulated through the temporal coupled-mode theory and can be derived from measured spectra such as transmissivity and reflectivity. We analyzed several metasurfaces supporting *quasi*-BIC numerically and experimentally to study the behavior of the spectral parameters as well as the resultant Purcell enhancement. We formulated the interdependence between the quality factor and the out-coupling efficiency, and revealed the existence of optimal detuning from the BIC. We also discovered that our findings are general and applicable towards realistic metasurfaces that are lossy and/or asymmetric. This discovery provides an intuitive model to understand the modal qualities of *quasi*-BIC and will facilitate optimization of *quasi*-BIC for luminescence enhancement applications.

## I. Introduction

Bound states in the continuum (BIC) is a concept that was first proposed in quantum mechanics, but eventually rose as a general wave phenomenon that has developed significant influence in numerous fields of physics.[1-5] In nanophotonics, BIC describe localized optical modes that share the same energy and momentum with planewaves (the continuum), but yet are incapable of coupling with the continuum, thus being a bound state.[6-8] There are two main types of BIC: symmetry-protected and accidental. Symmetry-protected BIC are prevented from coupling with planewaves due to symmetry mismatch between the localized optical modes and the planewave, while accidental BIC occur through Friedrich–Wintgen interference, where destructive interference between multiple leaky resonances cancels out each other and results in trapped states.[2,8-10] Due to the bound state nature, BIC exhibit diverging radiative quality ($Q$) factors, which theoretically can lead to an infinite $Q$ factor in an ideal system. In applications, BIC are often intentionally detuned to achieve weakly radiative modes with similarly high $Q$ factors, namely *quasi*-BIC ($q$-BIC).[11-16]

Photoluminescence on metasurfaces,[17-19] nanoplasmonics,[20-21] and photonic crystals[22-24] supporting BIC have been studied extensively due to the extraordinarily high $Q$ factor. The Purcell effect describes the enhancement in spontaneous emission rate of quantum emitters when placed in an optical cavity.[25-27] The increase in local density of optical states (LDOS) enhances the spontaneous emission rate, and the maximum enhancement is given by the Purcell factor $P_f = \frac{3}{4\pi^2}\left(\frac{\lambda}{n}\right)^3\frac{Q}{V}$, where $V$ is the effective modal volume.[25] As we look closer to the Purcell enhancement of $q$-BIC, we observe the unique feature that is the absence of any enhancement precisely at the BIC, even with precisely fabricated and high quality systems.[16-17,28] Intuitively, we understand this is due to the lack of out-coupling channels available for the emission as the detuning from the BIC diminishes. However, a comprehensive theory that accurately describes how this lack of out-coupling channels interacts with other factors that compose the Purcell factor, and ultimately leads to the absence of enhancement at the BIC, is still missing.

Recently, we derived an analytical model that uses spectral parameters to predict the average Purcell enhancement and examined the photoluminescence enhancement (PLE) in systems that support surface lattice resonance (SLR).[29] We discovered the spectral parameter that correlates with the nearfield confinement, which enables us to predict the Purcell enhancement while circumventing the need for measuring and integrating the nearfield. This ensures that our analytical model is compatible with experimental scenarios, where spectral measurements are reasonably accessible, but the nearfield is not. We also revealed that the three contributing factors to the measurable PLE are the $Q$ factor, the nearfield confinement, and the out-coupling efficiency, with the $Q$ factor being the most influential factor in SLR. Here, we expand upon this analytical model to analyze the influence of the out-coupling channels on the PLE of $q$-BIC. We first derive the equation that predicts the PLE of $q$-BIC with spectral parameters, in which we highlight the importance of the out-coupling efficiency for compatibility with $q$-BIC. The spectral parameters are defined based on the temporal coupled-mode theory (CMT) formalism, and the values are determined through fitting with measured or simulated spectra.[30-32] We strategically examine several metasurfaces outlined in Fig. 1 that support $q$-BIC through numerical and experimental analysis to gain a deeper understanding into the PLE. We first investigate a lossless scenario that resembles most closely an ideal symmetry-protected BIC and examine the basic spectral behaviors of $q$-BIC. Then, we step-by-step introduce material loss and random scattering loss into the analysis and discuss the effect of these realistic imperfections on Purcell enhancement. Finally, we also discuss the effects of asymmetry on the spectral properties and PLE of $q$-BIC through systems that support accidental $q$-BIC.

## II. Analytical Modelling of Purcell Enhancement of $q$-BIC

Numerous studies have shown that the enhancement in fractional LDOS can be predicted through the nearfield enhancement from a reciprocal source.[29,33-37] If we compare the emission that leaves the metasurface and propagates away as a planewave with momentum $-\hbar\mathbf{k}$ (and polarization $\mathbf{q}$), the fractional LDOS enhancement is proportional to the electric field intensity $|\mathbf{E}_\mathbf{k}(\mathbf{r},\omega)|^2$ under the incident $\hbar\mathbf{k}$ (and polarization $\mathbf{q}$), integrated over the space occupied by the emitters. Therefore, we can obtain the PLE for a specific emission angle by comparing the nearfield with a reference:

$$\mathrm{PLE}_{-\mathbf{k}}(\omega) = \frac{\int_{\mathrm{dye}} |\mathbf{E}_\mathbf{k}(\mathbf{r},\omega)|^2 \mathrm{d}^3 r}{\int_{\mathrm{dye}} \left|\mathbf{E}_{\mathrm{ref},\mathbf{k}}(\mathbf{r},\omega)\right|^2 \mathrm{d}^3 r} \tag{1}$$

where the $|\mathbf{E}_\mathbf{k}|^2$ of the resonance inside the dye medium under the incident $\hbar\mathbf{k}$ is compared to that of a reference structure $\left|\mathbf{E}_{\mathrm{ref},\mathbf{k}}\right|^2$ to calculate the enhancement in fractional radiative LDOS, and thus the PLE.

In our recent work, we discovered that the nearfield confinement of a resonance can be calibrated by the absorptive decay rate contributed by dye $\Gamma_{\mathrm{abs,dye}}$, where the nearfield region of interest is doped with a dye that absorbs at the resonant wavelength.[29,38] The absorptive decay rate $\Gamma_{\mathrm{abs}}$ is defined through the power absorbed $P_{\mathrm{abs}} = \Gamma_{\mathrm{abs}}|a(\omega_0)|^2$, where $a(\omega_0)$ is the mode amplitude in CMT and $|a(\omega_0)|^2$ is normalized to the total optical energy stored in the mode at resonant frequency $\omega_0$. The contribution of dye can then be identified by considering the power absorbed by the dye medium $P_{\mathrm{abs,dye}} = \Gamma_{\mathrm{abs,dye}}|a(\omega_0)|^2$. We can express $P_{\mathrm{abs,dye}}$ as the average dissipative energy density of the nearfield in the dye medium at $\omega_0$, giving us:

$$\Gamma_{\text{abs,dye}} = \frac{\int_{\text{dye}} \frac{1}{2}\omega_0 \varepsilon_0 \varepsilon''(\mathbf{r},\omega_0)|\mathbf{E}(\mathbf{r},\omega_0)|^2 \mathrm{d}^3 r}{|a(\omega_0)|^2}$$

(2)

where $\varepsilon_0$ is the permittivity of vacuum and $\varepsilon''$ is the imaginary part of the relative permittivity.[39-40] Under the assumption that the dye medium is homogeneous, we can treat $\varepsilon''(\mathbf{r},\omega_0)$ as a constant and factor it out of the integral, giving:

$$\int_{\text{dye}} |\mathbf{E}(\mathbf{r},\omega_0)|^2 \mathrm{d}^3 r = \frac{2|a(\omega_0)|^2 \Gamma_{\text{abs,dye}}}{\omega_0 \varepsilon_0 \varepsilon''_{\text{dye}}(\omega_0)}$$

(3)

which corresponds to the numerator in Eq. (1). Based on CMT, the steady-state solution gives $|a(\omega)|^2 = \frac{(\Gamma_{\text{rad}}/2)\sqrt{1-A_0}|\langle v^*|s_{+,\mathbf{k}}\rangle|^2}{(\omega-\omega_0)^2+(\Gamma_{\text{tot}}/2)^2}$, where $|v\rangle$ is the in-coupling constant and $|s_{+,\mathbf{k}}\rangle$ is the incident wave vector for the incident $\hbar\mathbf{k}$ in CMT.[41] $\Gamma_{\text{tot}} = \Gamma_{\text{rad}} + \Gamma_{\text{abs,dye}} + \Gamma_{\text{abs,MS}}$ is the total decay rate, $\Gamma_{\text{rad}}$ is the radiative decay rate and $\Gamma_{\text{abs,MS}}$ is the absorptive decay rate contributed by the metasurface (excluding the dye medium). Since we are primarily dealing with cases where the non-resonant absorptivity $A_0$ is zero at the emission wavelength, the $\sqrt{1-A_0}$ term equals 1.

For simplicity, we choose a reference that is a uniform thin film of the same dye-containing homogeneous medium, placed on the same substrate without the metasurface. Thus, we get a planar structure in which determining $\mathbf{E}_{\text{ref},\mathbf{k}}(\mathbf{r},\omega)$ becomes straightforward and we get $\langle s_{+,\mathbf{k}}|s_{+,\mathbf{k}}\rangle = \frac{1}{2}\sqrt{\frac{\varepsilon_0\varepsilon}{\mu_0\mu}}|\mathbf{E}_{\text{ref},\mathbf{k}}|^2 \times (\text{unit area})$. Combining the above equations gives:

$$\text{PLE}_{-\mathbf{k}}(\omega_0) = \frac{c\Gamma_{\text{rad}}}{\omega_0 t {\Gamma_{\text{tot}}}^2} \frac{\Gamma_{\text{abs,dye}}}{\kappa} \frac{|\langle v^*|s_{+,\mathbf{k}}\rangle|^2}{\langle s_{+,\mathbf{k}}|s_{+,\mathbf{k}}\rangle}$$

(4)

where $t = \int_{\text{dye}} 1/(\text{unit area})\, \mathrm{d}^3 r$ is the thickness of the reference dye medium, $\kappa$ is the extinction coefficient of the dye medium and $c$ is the speed of light in vacuum. We can slightly simplify Eq. (4) by defining the radiative decay rate associated with the incident $\hbar\mathbf{k}$ (and polarization $\mathbf{q}$) as $\Gamma_{\text{rad,in}} = \Gamma_{\text{rad}} \frac{|\langle v^*|s_{+,\mathbf{k}}\rangle|^2}{\langle s_{+,\mathbf{k}}|s_{+,\mathbf{k}}\rangle} = \Gamma_{\text{rad}}|v_{\mathbf{k}}^2|$ and dropping the subscript $-\mathbf{k}$, the PLE thus can be expressed as:

$$\text{PLE}(\omega_0) = \frac{c\Gamma_{\text{rad,in}}}{\omega_0 t {\Gamma_{\text{tot}}}^2} \frac{\Gamma_{\text{abs,dye}}}{\kappa}$$

(5)

## III. Results and Discussion

The influence of different factors on the PLE can be dissected by re-arranging Eq. (5) into $\mathrm{PLE}(\omega_0) = \left(\frac{\omega_0}{\Gamma_{\mathrm{tot}}}\right)\left(\frac{\Gamma_{\mathrm{abs,dye}}}{\kappa t}\right)\left(\frac{\Gamma_{\mathrm{rad,in}}}{\Gamma_{\mathrm{tot}}}\right)\left(\frac{c}{\omega_0^2}\right)$. The first term $\omega_0/\Gamma_{\mathrm{tot}}$ is the $Q$ factor. The second term $\Gamma_{\mathrm{abs,dye}}/(\kappa t)$ is related to $V$. As discussed in [29], $\Gamma_{\mathrm{abs,dye}}/(\kappa t)$ is proportional to the averaged $1/V$ over the dye medium in which the photoluminescence process occurs. The third term $\Gamma_{\mathrm{rad,in}}/\Gamma_{\mathrm{tot}}$ describes the out-coupling efficiency of the optical cavity, which is an influential factor when analyzing the PLE of $q$-BIC, yet was not captured in the classical representation of the Purcell factor.

**i. Lossless *q*-BIC on $TiO_2$ metasurface**

We can first get a basic understanding of the behavior of the PLE of $q$-BIC by considering an ideal, lossless metasurface that supports a symmetry-protected BIC. The detailed design of the metasurface is described in the Supplementary Materials.[41] The angle-dependent transmissivity $T$ and reflectivity $R$ of the metasurface were numerically simulated and plotted in Fig. 2(a) and 2(b). We can clearly observe the symmetry-protected BIC at incident angle $\theta = 0°$ and wavelength $\lambda =$ 612.6 nm, and as detuning $\theta$ was introduced, the $q$-BIC emerges with a $\pm\theta$ symmetry. The nearfield in the dye medium was also recorded and the numerically predicted PLE was calculated directly by Eq. (1) and plotted in Fig. 2(c). The $T$ and $R$ of each incident angle were fitted to obtain the spectral parameters $\omega_0$, $\Gamma_{\mathrm{tot}}$ and $\Gamma_{\mathrm{rad,in}}$, with Eq. (S4) and (S5) respectively.[41] The extinction coefficient $\kappa$ of the dye medium was modulated to calibrate the factor $\Gamma_{\mathrm{abs,dye}}/\kappa$, which is the slope of $\Gamma_{\mathrm{abs}}$ against $\kappa$.[41]

Since we focus on analyzing the behavior of the metasurface within a small detuning from the BIC, we can model the change in spectral parameters by a polynomial expansion, specifically up to the quadratic term. The $\omega_0$ and the decay rates are plotted in Fig. 2(d) and 2(e) with their polynomial fit. As illustrated in Fig. 2(e), $\Gamma_{\mathrm{abs}}$ remains zero for all incident angles while $\Gamma_{\mathrm{tot}}$ and $\Gamma_{\mathrm{rad,in}}$ increases quadratically from the BIC point. On the other hand, we can see that the factor $\Gamma_{\mathrm{abs,dye}}/\kappa$ remain mostly constant throughout the fitted range with a slight increase at larger $\theta$ (see Fig. S1(f)), which indicates similar nearfield confinement at different $\theta$. As a result, the total $Q$ factor diverges at the BIC and decreases following the inverse-squared law of the detuning ($\theta$).[42-44] We then predict the PLE with the best fit of the parameters and compare with the numerical PLE in Fig. 2(f). The PLE increases rapidly as we approach the BIC and appears to diverge at the BIC. Since non-radiative decay rates remain zero while $\Gamma_{\mathrm{tot}}$ and $\Gamma_{\mathrm{rad,in}}$ both varies quadratically against the incident angle, the out-coupling efficiency $\Gamma_{\mathrm{rad,in}}/\Gamma_{\mathrm{tot}}$ remains constant over all incident angles and did not influence (or reduce) the PLE.

However, as illustrated in Fig. 2(f), the behavior is significantly different once we artificially introduce a small $\Gamma_{\text{abs}}$, which models a non-zero absorption loss and/or random scattering loss. Since the $\Gamma_{\text{abs}}$ increases $\Gamma_{\text{tot}}$ slightly while not affecting $\Gamma_{\text{rad,in}}$, the out-coupling efficiency $\Gamma_{\text{rad,in}}/\Gamma_{\text{tot}}$ is no longer a constant but drops sharply around the BIC, influencing the PLE to hit a maximum and drop sharply around the BIC. We can see this effect even with very small artificial $\Gamma_{\text{abs}}$ equivalent to a linewidth broadening of 0.0001 nm, which is difficult to observe with conventional spectrometers. This prompts us to explore in more detail the effect of absorption and random scattering loss.

**ii. Lossy *q*-BIC on Si metasurface**

In the previous section, we revealed that even a very small absorption loss from the metasurface would inevitably change the characteristic behavior of PLE from BIC. However, material loss is inevitable in realistic metasurfaces;[45] nanofabrication also introduces inherent roughness into the metasurface structure, which introduces random scattering loss in application.[46] This prompts us to take a closer look into BIC supported on realistic materials with loss instead of an ideal, lossless model. Therefore, we consider a symmetry-protected BIC supported on a bipartite Si metasurface, as described in the Supplementary Materials.[41] The symmetry-protected BIC at $\lambda$ = 704.5 nm was detuned by displacing one of the Si nanoparticles by *d*, and the *d*-dependent transmissivity *T*, reflectivity *R*, absorptivity $A = 1 - T - R$ and PLE were numerically simulated and plotted in Fig. 3(a) – (d). The *T*, *R* and *A* were fitted to obtain the spectral parameters $\omega_0$, $\Gamma_{\text{tot}}$ and $\Gamma_{\text{rad,in}}$, while $\Gamma_{\text{abs,dye}}/\kappa$ was fitted by modulating $\kappa$ of the dye medium.[41]

As shown in Fig. 3(e), $\Gamma_{\text{abs}}$ remains almost constant against *d* with a small decrease at increasing *d*. This decrease is due to the gradual redshift of $\omega_0$ as *d* increases and Si being less absorptive at longer wavelengths. On the other hand, $\Gamma_{\text{rad,in}} = \Gamma_{\text{rad}}$ is zero at the BIC and increases quadratically from there. This is like what we observed in the lossless case, and we expect a quadratic dependency due to the limited range and the symmetry about ±*d*. Therefore, $\Gamma_{\text{tot}}$, the sum of $\Gamma_{\text{abs}}$ and $\Gamma_{\text{rad}}$, roughly follows the same shape of $\Gamma_{\text{rad}}$ but is displaced upwards by $\Gamma_{\text{abs}}$. Since *d* only shifts the position of the Si nanoparticles and does not increase or reduce the amount of dye medium, the factor $\Gamma_{\text{abs,dye}}/\kappa$ also remains mostly constant against *d*, as illustrated in Fig. S2(e).[41] This allows us to assume $\Gamma_{\text{abs,dye}}/\kappa$ is constant and use the value at *d* = 20 nm for the prediction of PLE. As illustrated in Fig. 3(f), the numerically simulated PLE aligned with the PLE predicted by Eq. (5). First, the PLE increases when reducing *d* and approaching the BIC due to the increase in *Q* factor. However, as we get closer to the BIC configuration, the decrease in the out-coupling efficiency overwhelms the improvement in *Q* factor and the PLE drops sharply near the BIC.

Interestingly, the maximum in PLE occurs very close to the point at which $\Gamma_{\text{abs}}$ and $\Gamma_{\text{rad}}$ intersects. This coincides with the critical coupling condition, which states that light-matter interaction is maximized when $\Gamma_{\text{abs}} = \Gamma_{\text{rad}}$.[32] This also presents a new perspective on how to optimize Purcell enhancement with realistic metasurfaces. First, $\Gamma_{\text{abs}}$ should be as small as possible, which can be achieved by using low-loss materials for the metasurface,[45] or reducing plasmonic loss with nonlocal resonances.[43,47] The $\Gamma_{\text{abs}}$ sets a lower limit to $\Gamma_{\text{tot}}$ and thereby also imposes an upper bound to the Q-factor. Then, $\Gamma_{\text{rad}}$ should be tuned to equal $\Gamma_{\text{abs}}$ to achieve the critical coupling condition and maximize light-matter interaction. If $\Gamma_{\text{rad}}$ is too small, the system is dominated by the absorptive loss and most of the enhanced emission would be absorbed by the metasurface instead of out-coupled. While for large $\Gamma_{\text{rad}}$, the increase in out-coupling efficiency is not worth the decrease in Q-factor and therefore the PLE is reduced.

### iii. Asymmetric $TiO_2$ metasurface supporting accidental *q*-BIC

In the following sections, we explore the behavior of an off-Γ accidental *q*-BIC on experimental and simulated metasurfaces. We fabricated asymmetric $TiO_2$ metasurfaces that support accidental *q*-BICs by glancing angle deposition of Ti followed by rapid thermal annealing (RTA) on symmetric $TiO_2$ metasurfaces.[41] A numerical model is built to replicate the behavior of the experimental asymmetric $TiO_2$ metasurface, which the design is described in the Supplementary Materials.[41] Since the $TiO_2$ was formed through RTA of Ti, we expect some roughness in the metasurface and the random scattering loss is modelled by adding an extinction coefficient $\kappa_{\text{TiO}_2} = 0.005$ to the $TiO_2$ material model. While the numerical simulation only simulates the coherent and periodic part of the EM field, the randomly scattered photons are decoherent from the resonant field. Therefore, we choose to model the random scattering loss as absorption as both absorbed photons and randomly scattered photons would not be part of the simulated field. We also assumed a uniform scattering density within the $TiO_2$ material, which is intended to account for both the surface roughness of the material as well as structural inhomogeneity within the material. As we can see in the angle-dependent *T*, *R*, *A* and PLE plotted in Fig. 4(a) – (d), an accidental BIC is observed at an oblique incident angle $\theta = 0.8°$ and $\lambda = 614$ nm. The *T*, *R* and *A* were fitted to obtain the spectral parameters $\omega_0$, $\Gamma_{\text{tot}}$ and $\Gamma_{\text{rad,in}}$, and $\Gamma_{\text{abs,dye}}/\kappa$ was also fitted.[41]

The decay rates of the accidental *q*-BIC on the asymmetric $TiO_2$ metasurface behaves markedly differently from that of the symmetric cases. As illustrated in Fig. 4(e), $\Gamma_{\text{abs}}$ is largest at normal incident and decreases following a linear style when increasing $|\theta|$. The $\Gamma_{\text{rad}}$ is also symmetrical over $\pm\theta$ and remains non-zero at all $\theta$. On the other hand, the asymmetry in the metasurface splits the point where $\Gamma_{\text{rad},\uparrow}$ and $\Gamma_{\text{rad},\downarrow}$ touch zero, corresponding to the ports at the superstrate (↑) and substrate (↓) sides, towards positive and negative incident angles respectively. $\Gamma_{\text{rad},\uparrow}$ and $\Gamma_{\text{rad},\downarrow}$ now touch zero at 0.92° and –0.53° respectively. To use the parameters to predict the PLE, the decay rates were fitted with fourth-order polynomials. The polynomial fits of the parameters were then substituted into Eq. (5) to derive a prediction of the PLE. As illustrated in Fig. 4(f), Eq. (5) predicted the numerical PLE with good accuracy (see Fig. S3(e) for the simulated substrate side PLE).[41] The PLE at the superstrate side shows a minimum at 0.9° while that at the substrate side shows a minimum at –0.5°. A local maximum in PLE is observed at –2.5° at the superstrate side and the model predicted it at –2.2°. On the substrate side, a local maximum is observed at 3.2° followed by a very slow decrease in PLE, while the model predicts a local maximum at 2.5°. Overall, we see larger discrepancies between the analytical and numerical PLE near the end of the fitted range. Due to the nature of polynomial regression, the edge of the data is usually less well represented compared to the data near the center (normal incident), as the higher-order terms are truncated and they are only significant for larger $|\theta|$. Therefore, in the trade-off between overfitting and poor fitting accuracy, we proceeded with fourth-order polynomials as they fitted the decay rates better when compared to second-order polynomials, while further increasing the polynomial order did not improve significantly upon that. While this choice is unlikely to lead us to overfitting, this also limits our accuracy, especially for the data near the edges of the fitted range.

It is worth noting that $R$ remained symmetric over $\pm\theta$, even on an asymmetric structure. We were also able to predict the PLE at both the superstrate side and the substrate side, only with the spectra incident from the superstrate side. Both features are because Lorentz reciprocity ensures the in-coupling and out-coupling process are symmetric, which ensures a symmetric $R$. Also, by simultaneously fitting $T$ and $R$ of $\pm\theta$, we were able to determine the in-coupling constants of all ports and thus model the PLE at both the superstrate side and the substrate side.

**iv. Asymmetric $TiO_2$ metasurface fabricated by glancing angle deposition**

In this section, we experimentally investigate the behavior of accidental $q$-BIC on the asymmetric $TiO_2$ metasurfaces. The asymmetric metasurface was fabricated by glancing angle deposition on a symmetric $TiO_2$ metasurface. A Ti thin film (100nm) was deposited on a $SiO_2$ glass substrate by electron beam deposition, and a resist (TU7, Obducat) was spin-coated on top. A resist array with square nanoparticles of sides 160 nm arranged in a square lattice of period 370 nm was then fabricated by nanoimprinting. The pattern was transferred to the Ti thin film by $O_2$ ashing (RIE-10NR) and etching with $Cl_2$ and $CHF_3$ (NLD-570). The Ti metasurface was then oxidized by RTA (900°C, 10 min) under $O_2$ atmosphere to fabricate a symmetric $TiO_2$ metasurface. A 20 nm thin film of Ti was deposited onto the $TiO_2$ metasurface from a glancing angle by tilting the substrate in an electron beam deposition apparatus. The substrate was tilted by 30° along the *xz*-plane. The newly deposited Ti was also oxidized with the same RTA process to create the asymmetric $TiO_2$ metasurface. As highlighted in the SEM image of the asymmetric metasurface shown in the inset of Fig. 5(d), voids formed only on the left-hand side of the $TiO_2$ nanoparticles, at the shadow of the nanoparticles during glancing angle deposition. Finally, a PMMA layer (450 nm) was spin-coated onto the $TiO_2$ metasurface. The PMMA is doped with an organic dye (Lumogen F 305 Red) at 1 wt.% (weight percent), and the PLE on this Lumogen dye is evaluated. The $T$, $R$ and PLE ($I/I_0$) were measured on the rotation stages illustrated in the Supplementary Materials.[41] The measured $T$, $R$ and PLE ($I/I_0$) are plotted in Fig. 5(a) – (c). We focus on the branch showing an accidental BIC at $\theta = 1.2°$ and $\lambda = 710$ nm. The spectral parameters $\omega_0$, $\Gamma_{\mathrm{tot}}$ and $\Gamma_{\mathrm{rad,in}}$ were fitted from the measured spectra directly. However, to calibrate the nearfield confinement with $\Gamma_{\mathrm{abs,dye}}/\kappa$, we need to use a dye that absorbs light at around $\lambda = 700$ – 800 nm. In place of the PMMA layer doped with Lumogen, we spin-coated a similar PMMA layer with 0.25 wt.% of IR780 iodide onto the asymmetric $TiO_2$ metasurface. The $T$ and $R$ of the metasurface with IR780 are then measured to fit for (mainly) $\Gamma_{\mathrm{abs}}$ and thus we derive $\Gamma_{\mathrm{abs,dye}}/\kappa$.[41]

The decay rates behaved like the simulated asymmetric $TiO_2$ metasurfaces. The $\Gamma_{\mathrm{abs}}$ is largest at normal incident and decreases slightly at larger $|\theta|$. As illustrated in Fig. 5(e), $\Gamma_{\mathrm{rad}}$ remains symmetric over $\pm\theta$, while $\Gamma_{\mathrm{rad},\uparrow}$ and $\Gamma_{\mathrm{rad},\downarrow}$ touches zero at 1.3° and –1.2° respectively. The parameters also show a good fit when approximated with fourth-order polynomials. We then use the polynomial fits to derive the predicted PLE through Eq. (5). While our analytical model only predicts enhancement due to the Purcell effect, enhancement in the absorption by the Lumogen dye due to the presence of the metasurface is not described, yet is included in the measured $I/I_0$. To minimize the influence of the absorption enhancement and the uneven excitation profile associated with higher absorption enhancement, the incident angle of the excitation laser was chosen at –31.5°, and the measured PLE was re-normalized against the baseline to isolate the Purcell effect. We can now compare the predicted PLE against the re-normalized PLE. As shown in Fig. 5(f), the experimental Purcell enhancement is predicted by our analytical model. We also apply Lorentz reciprocity to predict the backside PLE, which our experimental measurements show good agreement with (see Fig. S4(c) for the measured backside PLE).[41]

## IV. Conclusions

In conclusion, we leveraged the proposed analytical model to analyze the influence of the changes in the out-coupling channels of $q$-BIC on PLE. We first derived the equation that predicts the PLE of $q$-BIC with spectral parameters. Then, we strategically examined several metasurfaces that support $q$-BIC through numerical and experimental analysis to gain a deeper understanding of the PLE. We first investigated a lossless scenario that resembles most closely to an ideal symmetry-protected BIC and examined the basic spectral behavior of $q$-BIC. We then introduced material loss and random scattering loss into the analysis and discussed the effect of these realistic imperfections on the Purcell enhancement. Finally, we also discussed the effects of asymmetry on the spectral properties and PLE of $q$-BIC through systems that support accidental $q$-BIC. We revealed that the interplay between the quality factor and the out-coupling efficiency is the primary contribution towards the PLE of $q$-BIC and maximum enhancement can be achieved by optimizing the detuning from the associated BIC. We also demonstrated the general application methodology with two different dyes for emission measurement and absorption calibration respectively. This calibration approach with two optical materials can be further generalized to analyze different types of luminescent materials, making this analytical approach compatible even with those that do not inherently absorb light. Overall, this work provides an intuitive way to understand the mechanism of photoluminescence enhancement from $q$-BIC, as well as a practical framework for modelling the modal behavior of experimental $q$-BIC directly.

**RESEARCH FUNDING**

Financial support from Kakenhi (25K01501, 25K21709, 22H01776, 22K18884, 21H04619) is cordially acknowledged.

**AUTHOR CONTRIBUTION**

All authors have accepted responsibility for the entire content of this manuscript and consented to its submission to the journal, reviewed all the results and approved the final version of the manuscript. JTYT and SM conceived the idea. JTYT and TE designed and conducted the experiments. JTYT performed the simulations and analyzed the data. SM and KT supervised the study. JTYT prepared the manuscript with contributions from all co-authors.

**CONFLICT OF INTEREST**

Authors state no conflict of interest.

**DATA AVAILABILITY**

The data that support the findings of this article are available from the corresponding author upon reasonable request.

Figure 1. The metasurfaces that support $q$-BIC are illustrated. The numerically simulated (a) lossless $TiO_2$ metasurface, (b) lossy Si metasurface, (c) asymmetric $TiO_2$ metasurface and (d) experimental asymmetric $TiO_2$ metasurface are investigated in Section III (i) – (iv). The colors used are: $TiO_2$: green, $SiO_2$ (substrate): blue, Si: orange. The scale bar in the SEM image is 500 nm.

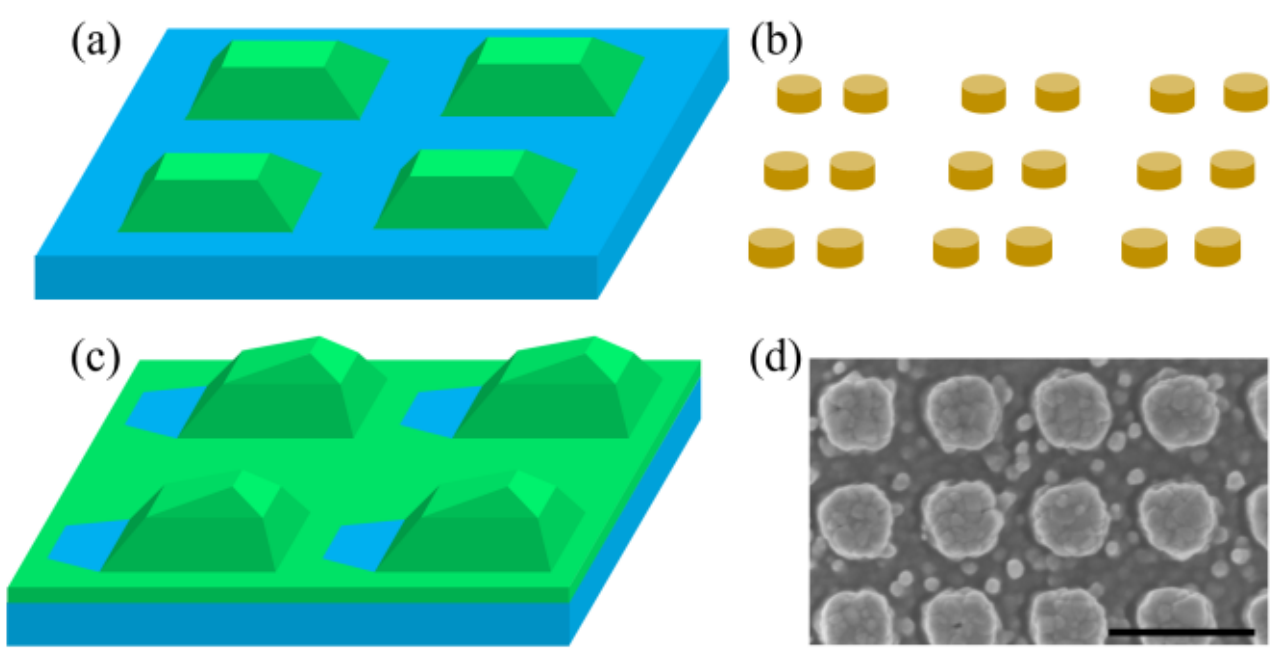

Figure 2. The numerically simulated (a) transmissivity, (b) reflectivity and (c) PLE of the lossless $TiO_2$ metasurface. The (d) resonant wavelength and (e) decay rates $\Gamma_{\text{tot}}$, $\Gamma_{\text{rad,in}}$ and $\Gamma_{\text{abs}}$ are plotted as a function of the incident angle $\theta$. (f) The numerical PLE and predicted PLE are plotted as a function of the emission angle $\theta$. The PLE predicted with a small artificial $\Gamma_{\text{abs}}$ are plotted as dashed lines.

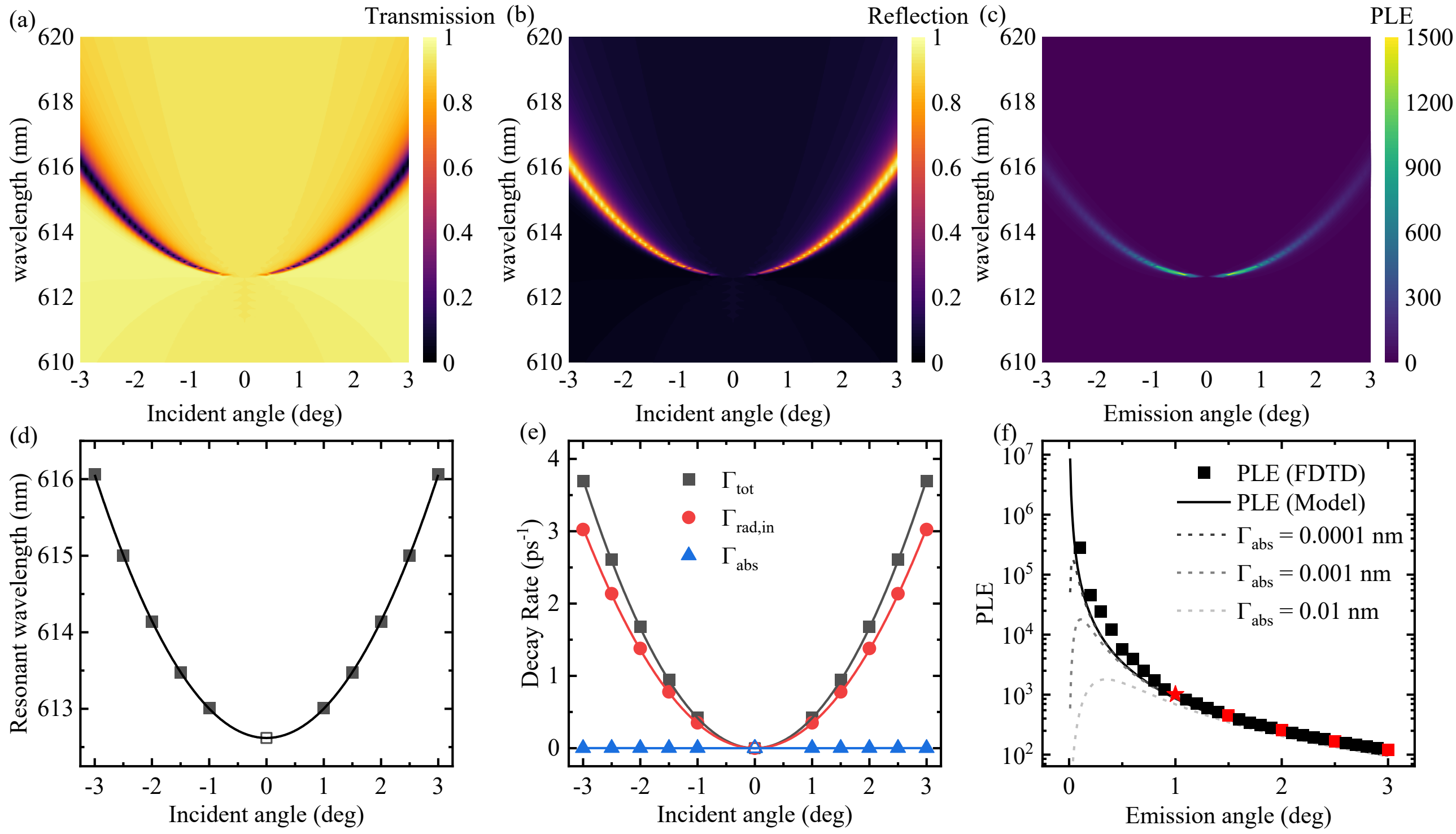

Figure 3. The numerically simulated (a) transmissivity, (b) reflectivity, (c) absorptivity and (d) PLE of the Si metasurface. (e) The decay rates $\Gamma_{tot}$, $\Gamma_{rad,in}$ and $\Gamma_{abs}$ are plotted as a function of the detuning *d*. (f) The numerical PLE and predicted PLE are plotted as a function of the detuning *d*.

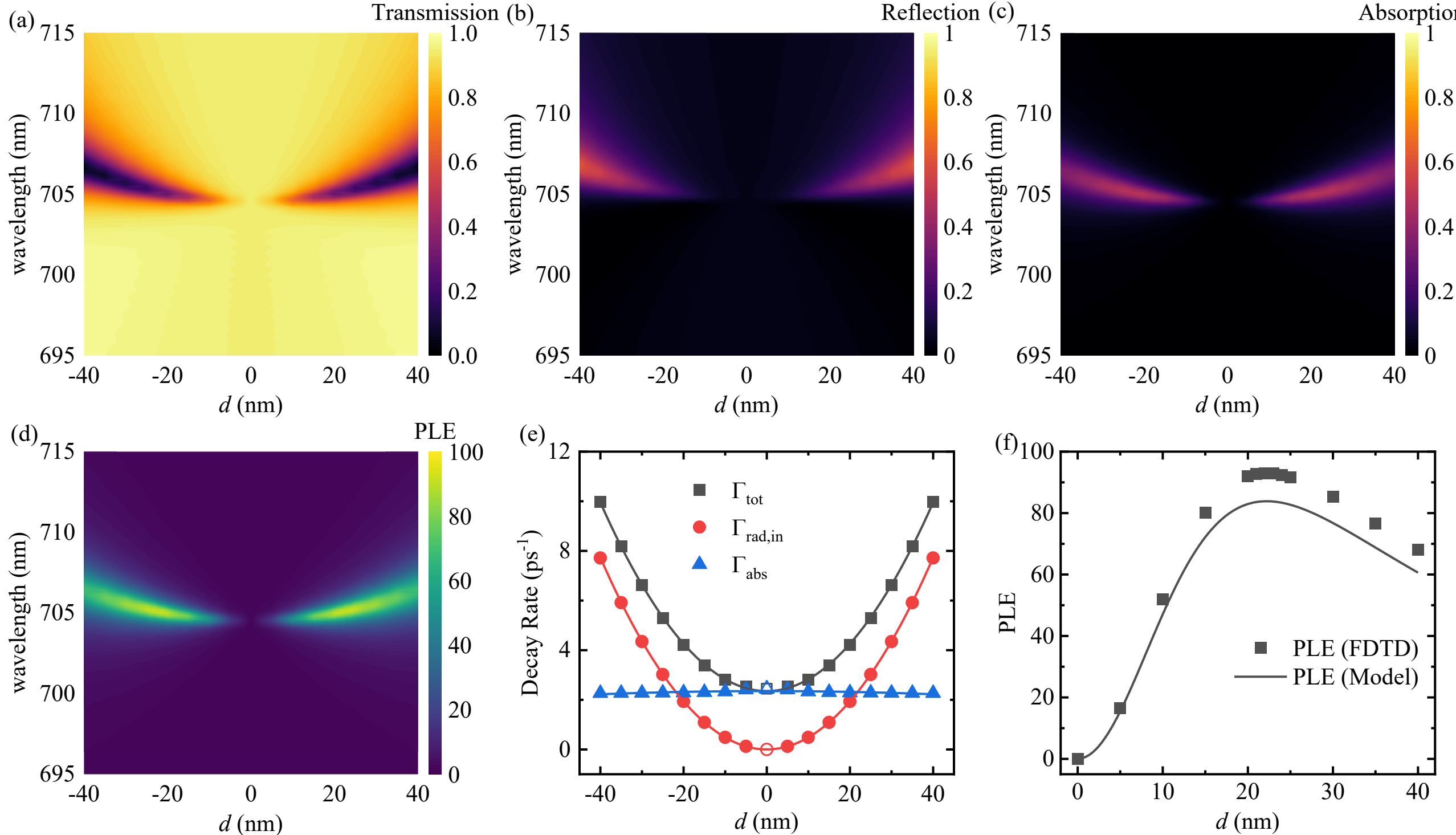

Figure 4. The numerically simulated (a) transmissivity, (b) reflectivity, (c) absorptivity and (d) PLE of the asymmetric $TiO_2$ metasurface. (e) The decay rates $\Gamma_{\mathrm{tot}}$, $\Gamma_{\mathrm{rad}}$, $\Gamma_{\mathrm{abs}}$, $\Gamma_{\mathrm{rad},\uparrow}$ and $\Gamma_{\mathrm{rad},\downarrow}$ are plotted as a function of the incident angle $\theta$. (f) The numerical PLE and predicted PLE are plotted as a function of the incident angle $\theta$.

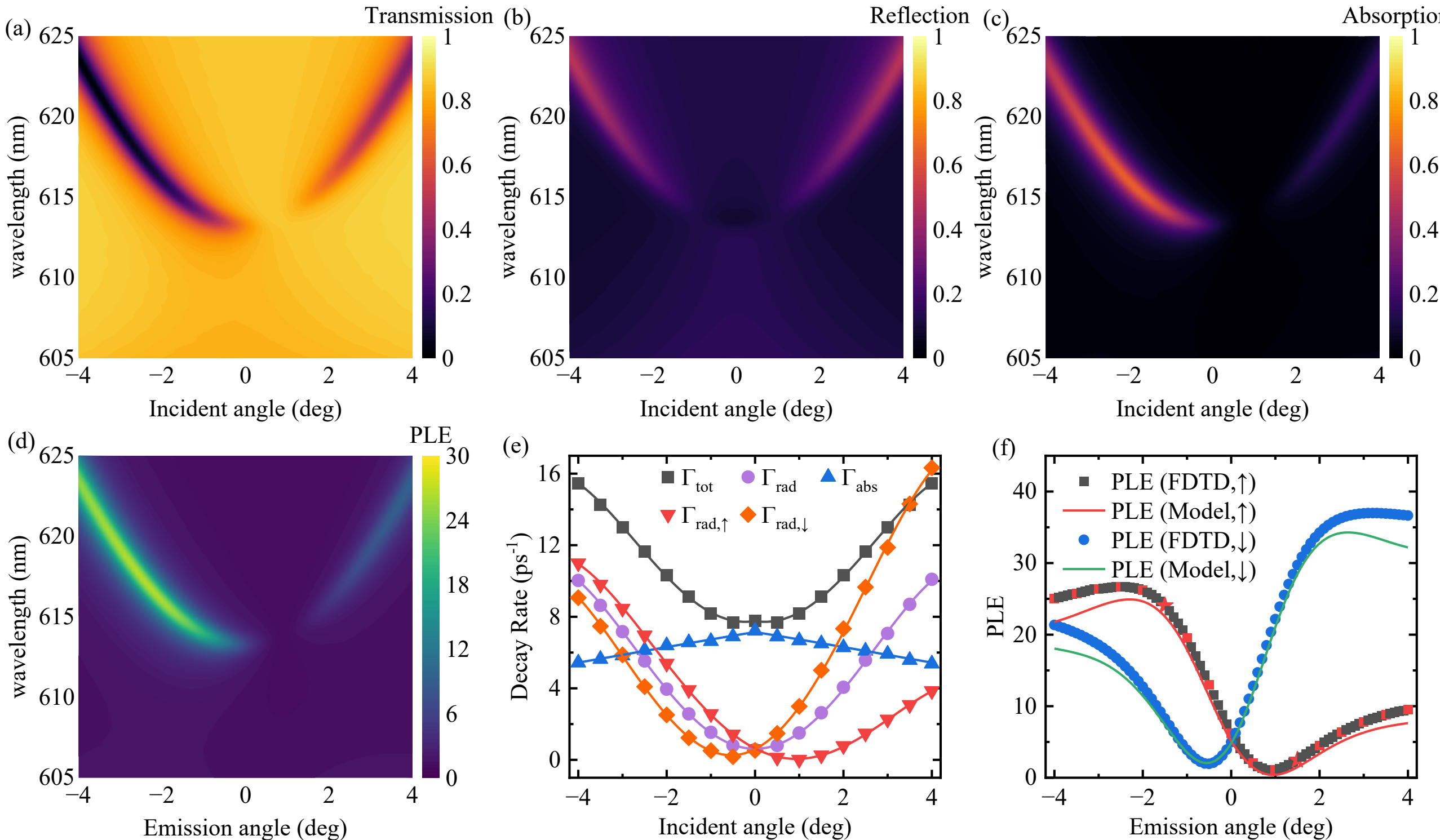

Figure 5. The measured (a) transmissivity, (b) reflectivity, and (c) PLE ($I/I_0$) of the asymmetric $TiO_2$ metasurface fabricated by glancing angle deposition. The arrows indicate the mode of interest. The (d) resonant wavelength and (e) decay rates $\Gamma_{\mathrm{tot}}$, $\Gamma_{\mathrm{rad}}$, $\Gamma_{\mathrm{abs}}$, $\Gamma_{\mathrm{rad},\uparrow}$ and $\Gamma_{\mathrm{rad},\downarrow}$ are plotted as a function of the incident angle $\theta$. The inset in (d) highlights the asymmetry in the $TiO_2$ metasurface. The void formed under the shadow of $TiO_2$ nanoparticles during glancing angle deposition are outlined by the dashed lines. (f) The re-normalized PLE and predicted PLE are plotted as a function of the emission angle $\theta$.

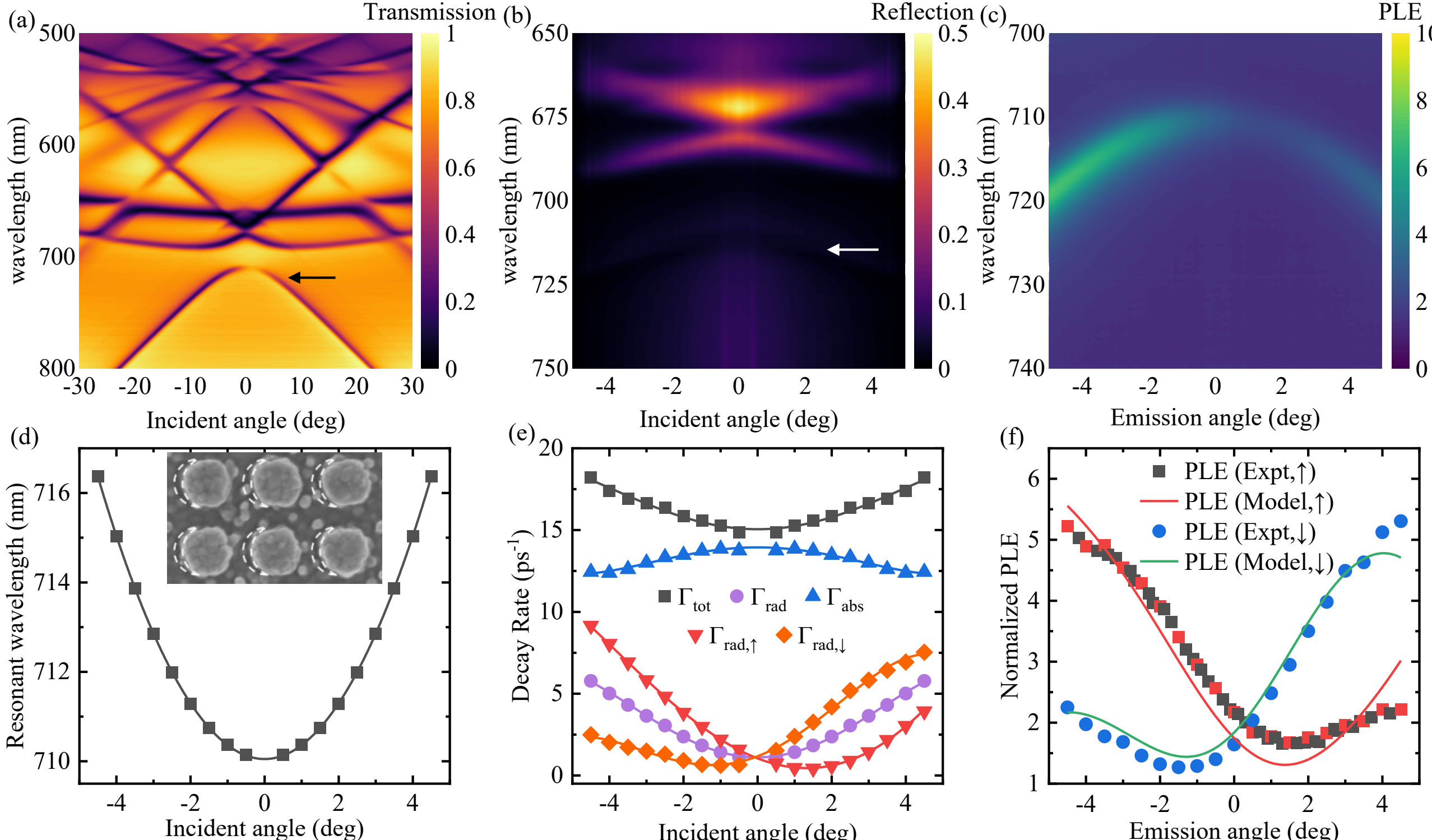

# Supplementary Materials: Modelling Purcell enhancement of metasurfaces supporting *quasi*-bound states in the continuum

Joshua T. Y. Tse[1,2]*, Taisuke Enomoto[1], Shunsuke Murai[2], and Katsuhisa Tanaka[1]

[1]Department of Material Chemistry, Graduate School of Engineering, Kyoto University, Katsura, Kyoto 615-8510, Japan

[2]Department of Physics and Electronics, Graduate School of Engineering, Osaka Metropolitan University, Osaka 599-8531, Japan

*Email: jtytse@omu.ac.jp

## S1. MODIFIED TEMPORAL COUPLED-MODE THEORY

We model the spectral behavior of the $q$-BIC modes with a modified CMT, which describes optical resonances as parametric oscillators and condenses the nearfield spatial profile into parameters that influence the spectral response of the resonators.[30-32] The mode amplitude $a$ of the $q$-BIC mode and the outgoing wave $|s_-\rangle$ follow the equations:

$$\frac{\mathrm{d}}{\mathrm{d}t}a(t) = \left(i\omega_0 - \frac{\Gamma_{\mathrm{tot}}}{2}\right)a(t) + \sqrt{\frac{\Gamma_{\mathrm{rad}}}{2}}\sqrt[4]{1-A_0}\langle v^*|s_+\rangle \tag{S1}$$

$$|s_-\rangle = \sqrt{1-A_0}\mathbf{C}|s_+\rangle + a(t)\sqrt{\frac{\Gamma_{\mathrm{rad}}}{2}}\sqrt[4]{1-A_0}|v\rangle \tag{S2}$$

where $\omega_0$ is the resonant frequency, $A_0$ is the non-resonant absorptivity and the total decay rate $\Gamma_{\mathrm{tot}}$ is the sum of the radiative decay rate $\Gamma_{\mathrm{rad}}$ and the absorptive decay rate $\Gamma_{\mathrm{abs}}$ of the $q$-BIC mode. $|s_\pm\rangle = (s_{\pm,\mathrm{TE},\uparrow} \quad s_{\pm,\mathrm{TM},\uparrow} \quad s_{\pm,\mathrm{TE},\downarrow} \quad s_{\pm,\mathrm{TM},\downarrow})^{\mathrm{T}}$ is the incident(+)/outgoing(–) wave vector normalized to the incident power,

$$\mathbf{C} = \begin{pmatrix} t_{0,\mathrm{TE}} & 0 & r_{0,\mathrm{TE}} & 0 \\ 0 & t_{0,\mathrm{TM}} & 0 & r_{0,\mathrm{TM}} \\ r_{0,\mathrm{TE}} & 0 & t_{0,\mathrm{TE}} & 0 \\ 0 & r_{0,\mathrm{TM}} & 0 & t_{0,\mathrm{TM}} \end{pmatrix}$$

is the direct scattering matrix where $t_{0,\mathrm{TE/TM}}$ and $r_{0,\mathrm{TE/TM}}$ are the non-resonant transmissivity and reflectivity coefficient respectively, and $|v\rangle = (v_{\mathrm{TE},\uparrow} \quad v_{\mathrm{TM},\uparrow} \quad v_{\mathrm{TE},\downarrow} \quad v_{\mathrm{TM},\downarrow})^{\mathrm{T}}$ is the in-coupling constant normalized by $\langle v|v\rangle = 2$. The subscript TE/TM denotes the polarization and ↑/↓ indicates the upwards/downwards ports at the superstrate/substrate sides respectively. We can solve for the steady-state solution of the modified CMT by considering solutions of form $a(t) = a(\omega)e^{i\omega t}$, which gives:

$$i\omega a(\omega) = \left(i\omega_0 - \frac{\Gamma_{\mathrm{tot}}}{2}\right)a(\omega) + \sqrt{\frac{\Gamma_{\mathrm{rad}}}{2}}\sqrt[4]{1-A_0}\langle v^*|s_+\rangle \tag{S3}$$

from Eq. (S1) and can be simplified to $a(\omega) = \frac{\sqrt{\Gamma_{\mathrm{rad}}/2}\sqrt[4]{1-A_0}\langle v^*|s_+\rangle}{i(\omega-\omega_0)+\Gamma_{\mathrm{tot}}/2}$ or $|a(\omega)|^2 = \frac{(\Gamma_{\mathrm{rad}}/2)\sqrt{1-A_0}|\langle v^*|s_+\rangle|^2}{(\omega-\omega_0)^2+(\Gamma_{\mathrm{tot}}/2)^2}$. Then, we can use the steady-state solution to derive the (TE-) transmissivity and reflectivity with incident from the superstrate side to be:

$$T_{\mathrm{TE}} = (1-A_0)\left|t_{0,\mathrm{TE}} + \frac{\Gamma_{\mathrm{rad}}}{2}\frac{v_{\mathrm{TE},\uparrow(\mathrm{in})}v_{\mathrm{TE},\downarrow(\mathrm{out})}}{i(\omega-\omega_0)+\Gamma_{\mathrm{tot}}/2}\right|^2 \tag{S4}$$

$$R_{\mathrm{TE}} = (1-A_0)\left|r_{0,\mathrm{TE}} + \frac{\Gamma_{\mathrm{rad}}}{2}\frac{v_{\mathrm{TE},\uparrow(\mathrm{in})}v_{\mathrm{TE},\uparrow(\mathrm{out})}}{i(\omega-\omega_0)+\Gamma_{\mathrm{tot}}/2}\right|^2 \tag{S5}$$

The subscript in/out denotes whether the coupling constant is associated with the incident or the outgoing port. This distinction is helpful when dealing with oblique incidents on asymmetric metasurfaces as they use different ports (with different coupling constants) for in-coupling and out-coupling. The absorptivity under the same incident can also be derived by considering $A = 1 - \frac{\langle s_-|s_-\rangle}{\langle s_+|s_+\rangle}$, which we find:

$$A_{\mathrm{TE}} = A_0 + (1-A_0)\frac{\Gamma_{\mathrm{abs}}\Gamma_{\mathrm{rad}}}{2}\frac{\left|v^2_{\mathrm{TE},\uparrow(\mathrm{in})}\right|}{(\omega-\omega_0)^2+(\Gamma_{\mathrm{tot}}/2)^2} \tag{S6}$$

where $\Gamma_{\mathrm{rad}}\left|v^2_{\mathrm{TE},\uparrow(\mathrm{in})}\right| = \Gamma_{\mathrm{rad}}\frac{|\langle v^*|s_+\rangle|^2}{\langle s_+|s_+\rangle} \equiv \Gamma_{\mathrm{rad,in}}$. Similar results can also be derived for the TM counterpart, where the subscripts TE are replaced by TM. More importantly, the spectral parameters used in Eq. (5), $\omega_0$, $\Gamma_{\mathrm{tot}}$ and $\Gamma_{\mathrm{rad,in}}$, can be determined by fitting the measured/simulated spectra with Eq. (S4) – (S6).

## S2. NUMERICAL SIMULATION DETAILS

We used finite-difference time-domain method (Ansys Lumerical FDTD) to numerically simulate the properties of metasurfaces that support BIC. In Section III (i), we considered an ideal, lossless metasurface that supports symmetry-protected BIC, as illustrated in the inset of Fig. S1(f). The metasurface consists of $TiO_2$ (refractive index $n$ = 2.4) nanoparticle array placed on a $SiO_2$ ($n$ = 1.46) substrate and covered by a PMMA ($n$ = 1.49) thin film. The $TiO_2$ nanoparticles are frustums with a rectangular base of 240 nm by 200 nm, top of 180 nm by 160 nm and height 160 nm, arranged in a 2D square lattice of period 370 nm. The PMMA thin film is 300 nm thick and acts as the dye medium which the nearfield is integrated to compute the numerical PLE. (An extinction coefficient $\kappa$ is added to the PMMA when fitting for $\Gamma_{\mathrm{abs,dye}}/\kappa$, see Section S3) The $SiO_2$ substrate and the vacuum above the PMMA layer are semi-infinite and extend beyond the simulation domain. The $x$- and $y$-boundaries are terminated with periodic boundary conditions to simulate an infinitely extending metasurface, and the $z$-boundaries are terminated by perfectly matched layers (PML).

In Section III (ii), we numerically simulated the symmetry-protected BIC supported on a bipartite Si metasurface, as illustrated in the inset of Fig. S2(e). The metasurface consists of two Si nanocylinder suspended in an index-matching layer ($n$ = 1.46) of thickness 200 nm, and sandwiched by two semi-infinite $SiO_2$ ($n$ = 1.46) layers. The Si nanocylinders with diameter 130 nm and height 90 nm are arranged in a 2D rectangular lattice of periodicity $P_x$ = 470 nm and $P_y$ = 235 nm. The complex refractive index of Si is extracted from [48]. The nanocylinders are aligned in $y$ and separated by $P_x/2 - d$ in $x$. The center of the index-matching layer is vertically aligned with the Si nanocylinders, and it also acts as the dye medium where the nearfield is integrated to compute the numerical PLE. (An extinction coefficient $\kappa$ is added to the index-matching layer when fitting for $\Gamma_{\mathrm{abs,dye}}/\kappa$.) Like the previous one, the $x$- and $y$-boundaries are terminated with periodic boundary conditions, and the $z$-boundaries are terminated by PMLs.

In Section III (iii), we built a numerical model that replicates the behavior of an asymmetric $TiO_2$ metasurface. The asymmetric $TiO_2$ metasurface we referred to was fabricated following the procedures described in Section III (iv), but with the glancing angle deposition done at 45°. The design of the metasurface is illustrated in Fig. S3(f). The designed metasurface consists of a $TiO_2$ (refractive index $n = 2.4 + 0.005i$) structure placed on a $SiO_2$ ($n$ = 1.46) substrate and covered by a PMMA ($n$ = 1.49) thin film. The $TiO_2$ metasurface consists of a slanted frustum with a triangular top, laying on a film of $TiO_2$ with a hole at the shadow of the nanostructure. The PMMA thin film is up to 300 nm thick and acts as the dye medium which the nearfield is integrated to compute the numerical PLE. (An extinction coefficient $\kappa$ is added to the PMMA when fitting for $\Gamma_{\mathrm{abs,dye}}/\kappa$.) Like in Section

III (i), the $SiO_2$ substrate and the vacuum above the PMMA layer are semi-infinite and extend beyond the simulation domain. The *x*- and *y*-boundaries are also terminated with periodic boundary conditions, and the *z*-boundaries are terminated by PMLs.

## S3. THE NEARFIELD CONFINEMENT FACTOR $\Gamma_{\mathrm{abs,dye}}/\kappa$

In Section III (i), we studied an ideal, lossless metasurface that supports symmetry-protected BIC. The nearfield confinement of the *q*-BIC is parametrized by the factor $\Gamma_{\mathrm{abs,dye}}/\kappa$ as we introduce an extinction coefficient $\kappa$ into the dye medium. Due to the high $Q$ factor, we use a relatively small $\kappa$ to minimize the disturbance towards the nearfield distribution of the *q*-BIC mode. We numerically simulated the $T$ and $R$ for $\theta = 1°$, 1.5°, 2°, 2.5°, 3°. $\kappa$ was varied from 0.00002 to 0.0001 for $\theta = 1°$ and from 0.0001 to 0.0005 for the rest. The evolution of $\Gamma_{\mathrm{tot}}$, $\Gamma_{\mathrm{rad}}$ and $\Gamma_{\mathrm{abs}}$ against $\kappa$ are shown in Fig. S1(a) – (e). $\Gamma_{\mathrm{abs,dye}}/\kappa$ is extracted from the slope of $\Gamma_{\mathrm{abs}}$ against $\kappa$ and is plotted in Fig. S1(f) for different $\theta$.

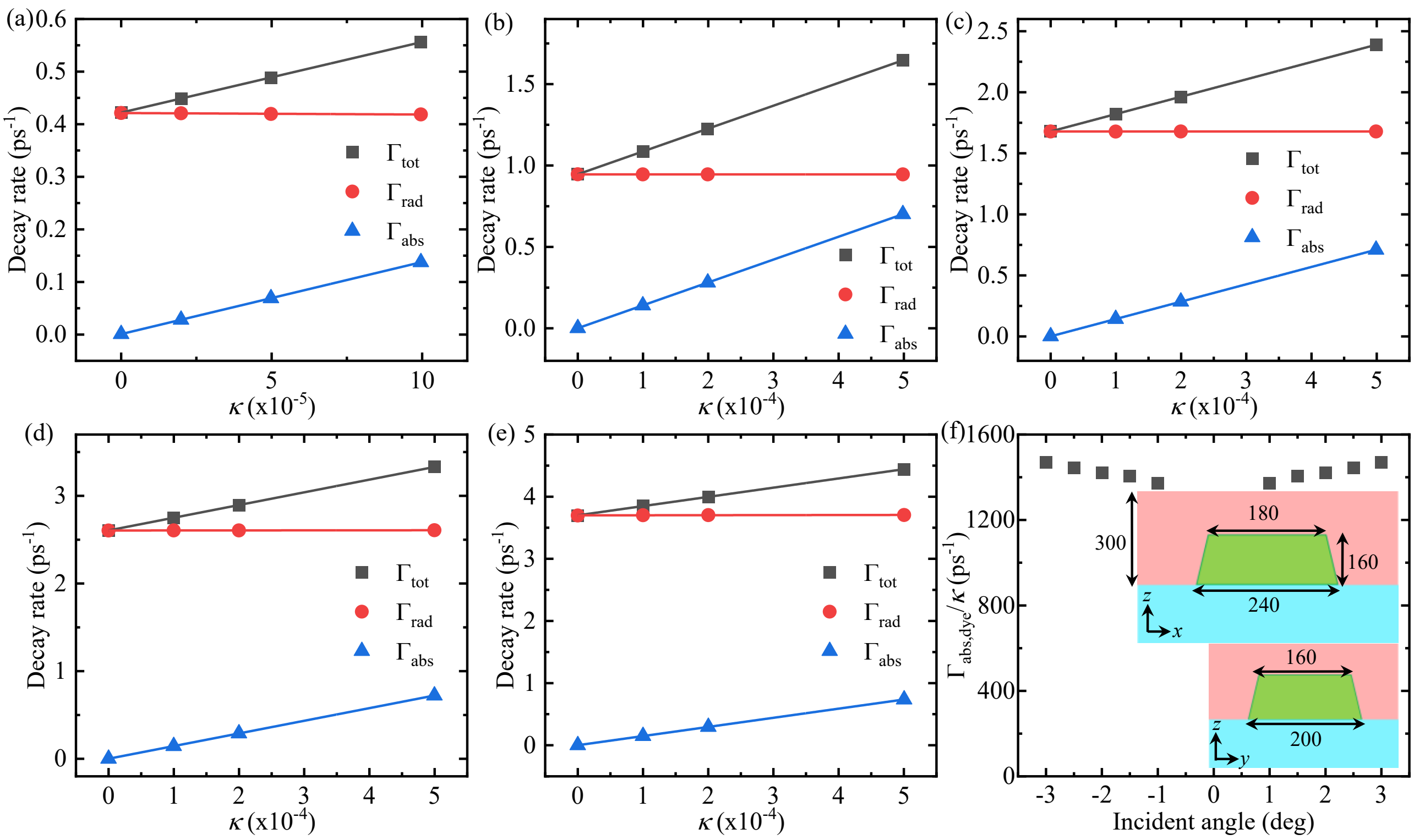

Figure S1. The decay rates $\Gamma_{\mathrm{tot}}$, $\Gamma_{\mathrm{rad}}$, and $\Gamma_{\mathrm{abs}}$ of the lossless $TiO_2$ metasurface at $\theta$ = (a) 1°, (b) 1.5°, (c) 2°, (d) 2.5°, and (e) 3° are plotted as a function of the artificial extinction coefficient $\kappa$ of the dye medium. (f) The resultant $\Gamma_{\mathrm{abs,dye}}/\kappa$ is plotted as a function of $\theta$. The inset shows the *xz*- and *yz*-cross section of the metasurface. The pink layer represents the PMMA with dye, the blue layer represents the $SiO_2$ substrate, and the green trapezoid represents the $TiO_2$ nanoparticle. The dimensions are given in nm.

In Section III (ii), we studied a bipartite Si metasurface that supports lossy $q$-BIC. The nearfield confinement of the $q$-BIC is parametrized by the factor $\Gamma_{\mathrm{abs,dye}}/\kappa$ as we introduce an extinction coefficient $\kappa$ into the dye medium. We numerically simulated the $T$, $R$ and $A$ for $d$ = 10, 20, 30, 40 nm. $\kappa$ was varied from 0.0025 to 0.01. The evolution of $\Gamma_{\mathrm{tot}}$, $\Gamma_{\mathrm{rad}}$ and $\Gamma_{\mathrm{abs}}$ against $\kappa$ are shown in Fig. S2(a) – (d). $\Gamma_{\mathrm{abs,dye}}/\kappa$ is extracted from the slope of $\Gamma_{\mathrm{abs}}$ against $\kappa$ and is plotted in Fig. S2(e) for different $d$.

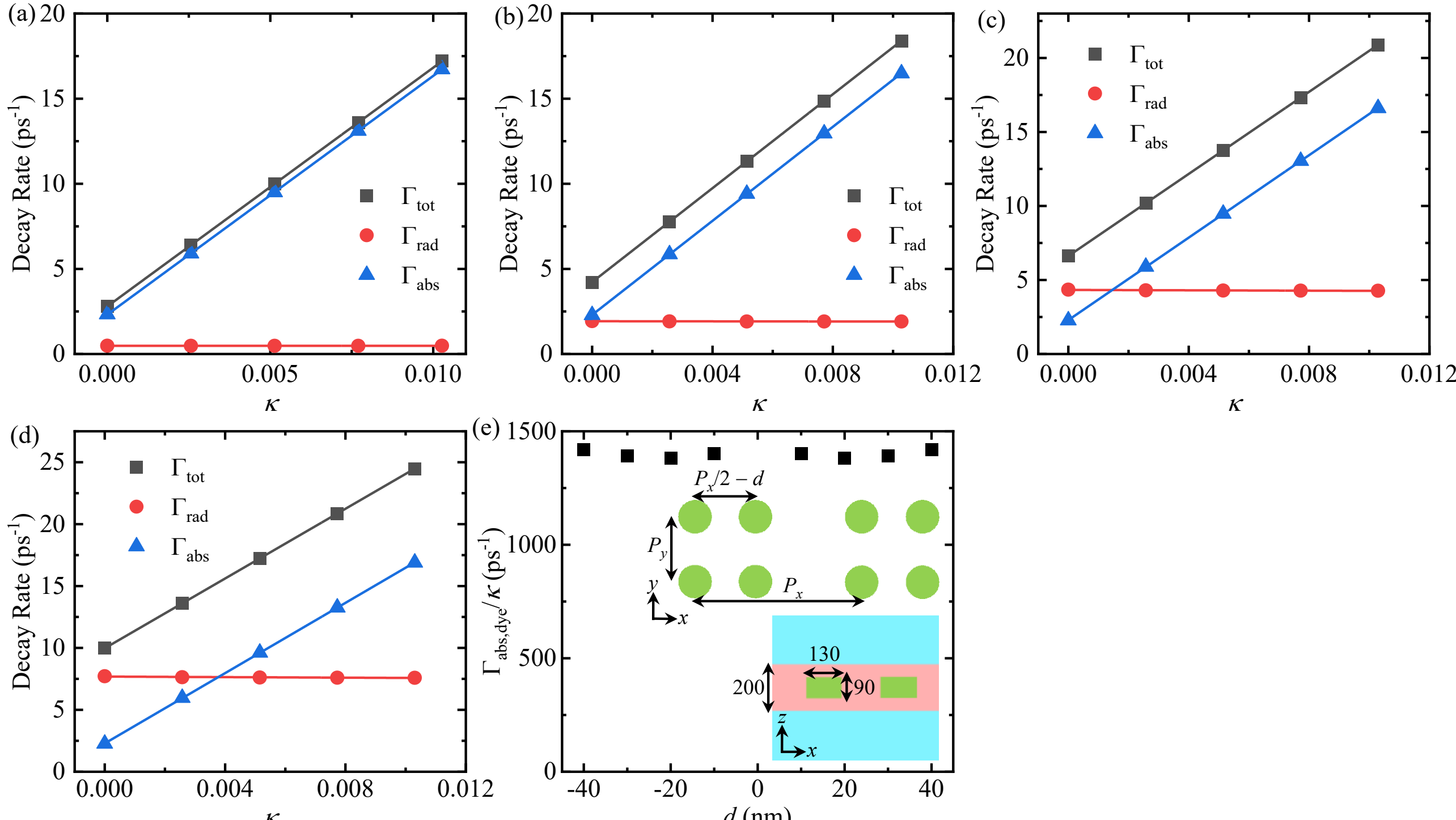

Figure S2. The decay rates $\Gamma_{\mathrm{tot}}$, $\Gamma_{\mathrm{rad}}$, and $\Gamma_{\mathrm{abs}}$ of the Si metasurface at $d$ = (a) 10, (b) 20, (c) 30, and (d) 40 nm are plotted as a function of the artificial extinction coefficient $\kappa$ of the dye medium. (e) The resultant $\Gamma_{\mathrm{abs,dye}}/\kappa$ is plotted as a function of $d$. The inset shows the $xy$- and $xz$-cross section of the Si metasurface. The pink layer represents the index-matching layer with dye, the blue layers represent the $SiO_2$ substrate and superstrate, and the green circles (rectangles) represent the Si nanoparticles. The dimensions are given in nm.

In Section III (iii), we studied an asymmetric $TiO_2$ metasurface that supports accidental $q$-BIC. The nearfield confinement of the $q$-BIC is parametrized by the factor $\Gamma_{\mathrm{abs,dye}}/\kappa$ as we introduce an extinction coefficient $\kappa$ into the dye medium. We numerically simulated the $T$, $R$ and $A$ for $\theta = \pm 0.5°$, $\pm 1.5°$, $\pm 2.5°$. $\kappa$ was varied from 0.005 to 0.01. The evolution of $\Gamma_{\mathrm{tot}}$, $\Gamma_{\mathrm{rad}}$ and $\Gamma_{\mathrm{abs}}$ against $\kappa$ are shown in Fig. S3(a) – (c). $\Gamma_{\mathrm{abs,dye}}/\kappa$ is extracted from the slope of $\Gamma_{\mathrm{abs}}$ against $\kappa$ and is plotted in Fig. S3(d) for different $\theta$.

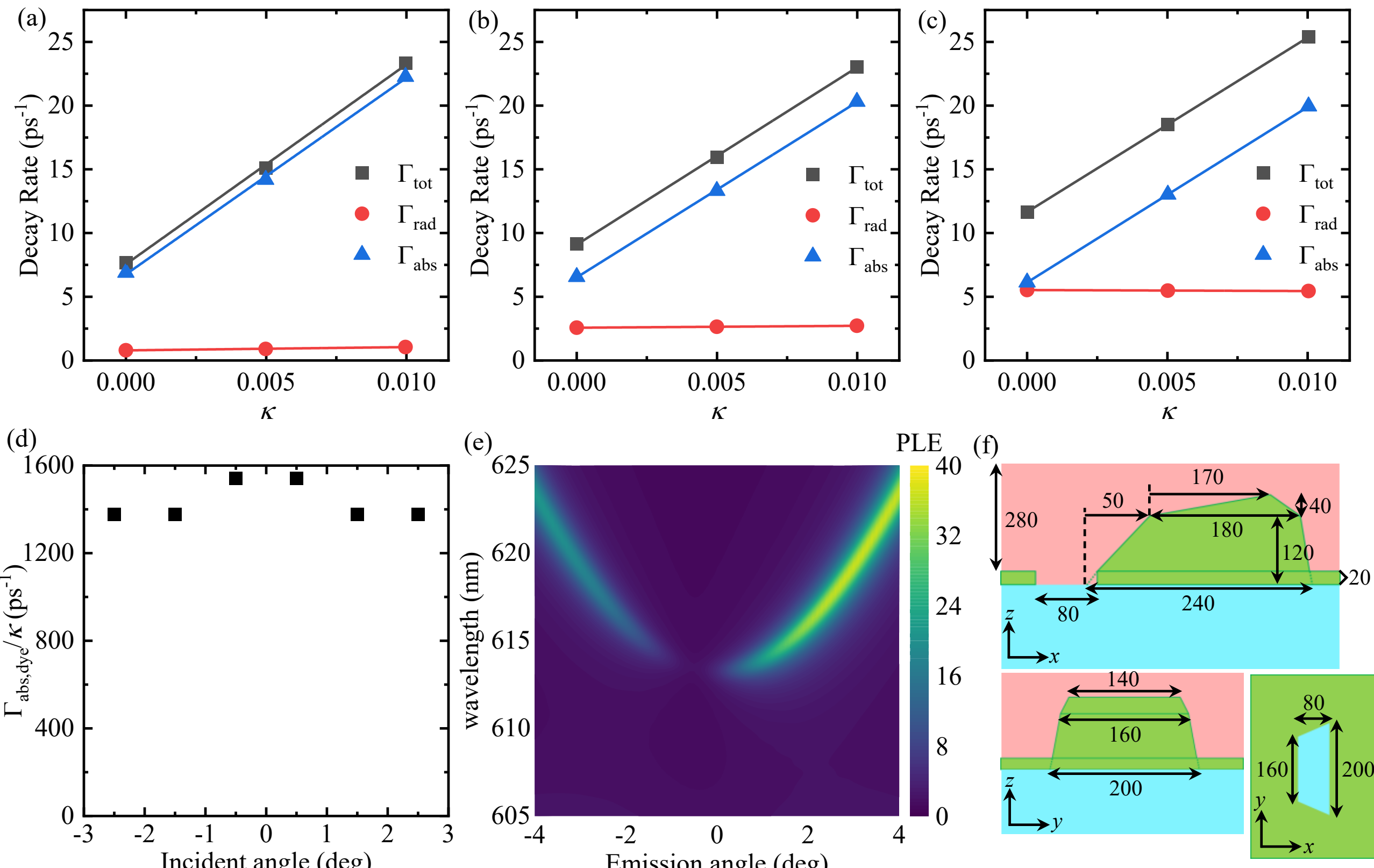

Figure S3. The decay rates $\Gamma_{\text{tot}}$, $\Gamma_{\text{rad}}$, and $\Gamma_{\text{abs}}$ of the asymmetric $TiO_2$ metasurface at $\theta$ = (a) ±0.5°, (b) ±1.5°, and (c) ±2.5° are plotted as a function of the artificial extinction coefficient $\kappa$ of the dye medium. (d) The resultant $\Gamma_{\text{abs,dye}}/\kappa$ is plotted as a function of $\theta$. (e) The numerically simulated substrate side PLE of the asymmetric $TiO_2$ metasurface. (f) The *xz*- and *yz*-cross section of the asymmetric $TiO_2$ metasurface is illustrated. The *xy*-cross section of the hole in the $TiO_2$ is also illustrated. The pink layer represents the PMMA with dye, the blue layer represents the $SiO_2$ substrate, and the green parts represent the $TiO_2$ metasurface. The dimensions are given in nm.

In Section III (iv), we experimentally studied an asymmetric $TiO_2$ metasurface that supports accidental *q*-BIC. The nearfield confinement of the *q*-BIC is parametrized by the factor $\Gamma_{\text{abs,dye}}/\kappa$ as we introduce the IR780 iodide dye into the PMMA layer. A PMMA layer doped with 0.25 wt.% of IR780 iodide was spin-coated on the asymmetric $TiO_2$ metasurface in place of the PMMA layer doped with the Lumogen dye. The *T* and *R* are measured and plotted in Fig. S4(a) and S4(b). The *T* and *R* are fitted for $\theta$ = ±1°, ±2°, ±3°, ±4°, and the evolution of $\Gamma_{\text{abs}}$ against $\kappa$ is shown in Fig. S4(d). (The value of $\kappa$ is discussed in Section S5.) $\Gamma_{\text{abs,dye}}/\kappa$ is extracted from the slope of $\Gamma_{\text{abs}}$ against $\kappa$.

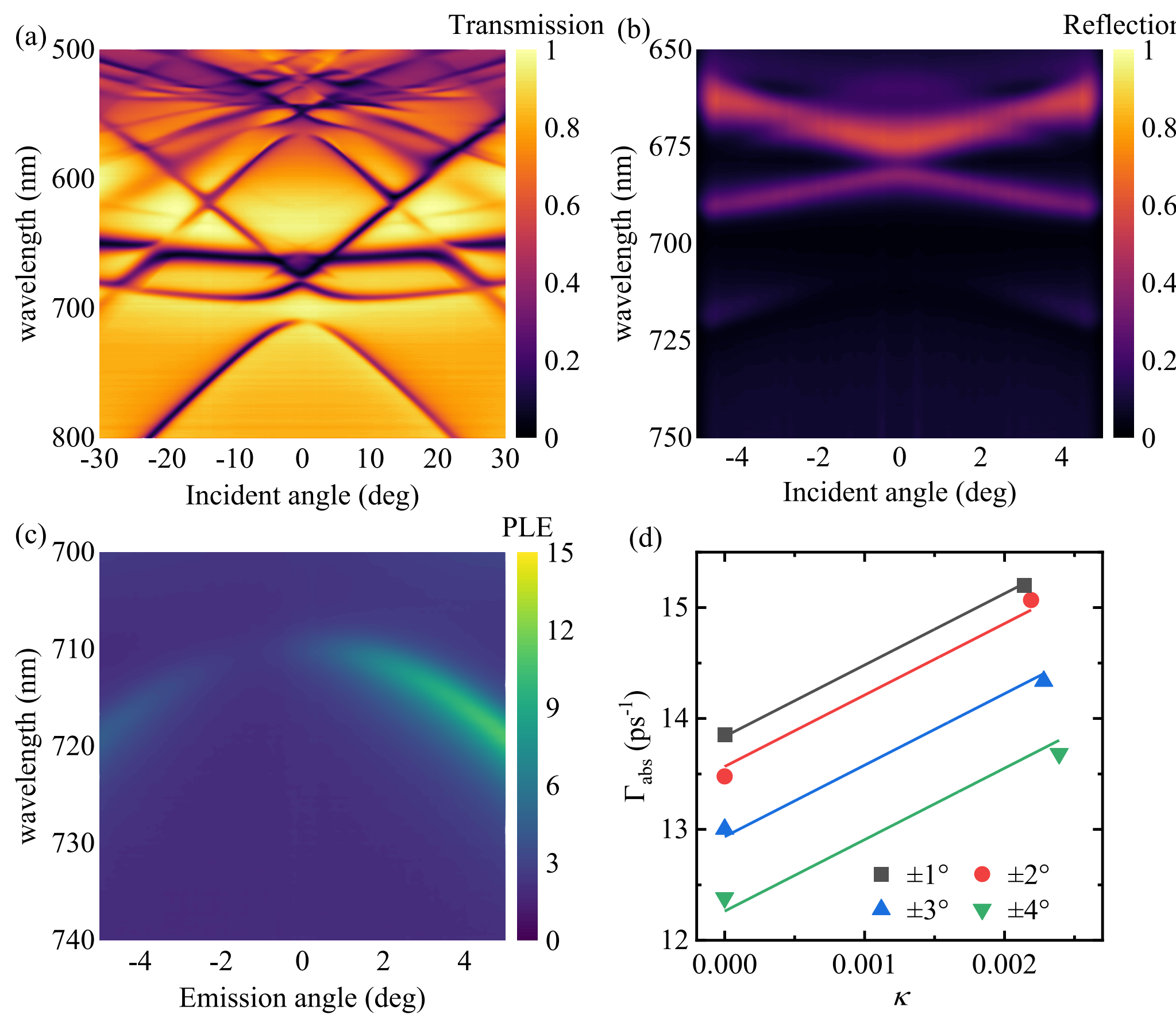

Figure S4. The measured (a) transmissivity and (b) reflectivity of the asymmetric $TiO_2$ metasurface spin-coated with IR780-containing PMMA. (c) The measured substrate side PLE ($I/I_0$) of the asymmetric $TiO_2$ metasurface (with Lumogen). (d) The $\Gamma_{abs}$ at $\theta = \pm 1°, \pm 2°, \pm 3°, \pm 4°$are plotted as a function of $\kappa$.

## S4. OPTICAL MEASUREMENT SETUPS

The $T$, $R$ and PLE of metasurfaces are measured optically. The schematic of the measurement setup for transmissivity measurements is shown in Fig. S5(a). A stabilized Tungsten-Halogen lamp (Thorlabs SLS201L) is used as a white light source. The light is coupled through a multimode fiber (Thorlabs M20L01) and collimated with a 5X microscope objective (Nikon MUE42050). The polarization of the incident light is controlled by a polarizer (Thorlabs WP25M-VIS). The light is then directed to the metasurface sample from the PMMA (superstrate) side. The sample is placed on a rotation stage, and the incident angle is controlled by rotating the sample. The transmitted light is collected through a lens (Thorlabs LB1901) into a multimode fiber (Ocean Optics P400-2-UV-VIS) connected to a spectrometer (Ocean Optics FLAME-S).

The schematic of the measurement setup for reflectivity measurements is shown in Fig. S5(b). A Tungsten-Halogen lamp (Ocean Optics HL-2000) is used as a white light source. The light is coupled through a multimode fiber (Ocean Optics QP400-2-UV-VIS) and collimated with a 5X microscope objective (Nikon MUE42050). The polarization of the incident light is controlled by a polarizer (Thorlabs WP25M-VIS). The light is then directed through a cube beamsplitter (Thorlabs CCM1-BS013) to the metasurface sample from the PMMA (superstrate) side. The reflected light is first reflected by the beamsplitter, and then collected through a lens (Thorlabs LB1901) into a multimode fiber (Ocean Optics P400-2-UV-VIS) connected to a spectrometer (Ocean Optics FLAME-S). The incident arm and the detector arm are mounted on rotation stages and rotate in opposite directions to measure the angle-dependent reflectivity. Due to the constraint from the aperture size of the beamsplitter mount, the reflectivity can only be measured within a ~6° window. However, as discussed in Section III (iii), the reflectivity is expected to be symmetric over $\pm\theta$ due to Lorentz reciprocity. Therefore, the range of the reflection setup is adjusted to utilize larger incident angle rather than to be symmetric about normal incident, resulting in a working range of $-1°$ to 5°.

The experimental setup for emission measurement is shown in Fig. S5(c). A 445 nm blue laser (SLOC BLS-445-1000mW) is used as the excitation source of the Lumogen dye from the substrate side. The collimated laser beam is controlled to be *p*-polarized by a Glan-Thompson polarizer (OptoSigma GTPB-08-21SN) and the incident angle is fixed at $\theta_{in} = 31.5°$. The emission polarization is filtered by a linear polarizer (Thorlabs WP25M-VIS) and the emission $I$ at the superstrate side and angle $\theta$ is focused by a lens (Thorlabs LB1901) into an optical fiber (Ocean Optics P400-2-UV-VIS) connected to a spectrometer (Ocean Optics FLAME-S). The detection arm is rotated around the sample to measure the emission at different $\theta$. The emission from the metasurface sample is normalized against the emission intensity of the Lumogen dye on an unstructured layer $I_0$ to obtain the PLE ($I/I_0$).

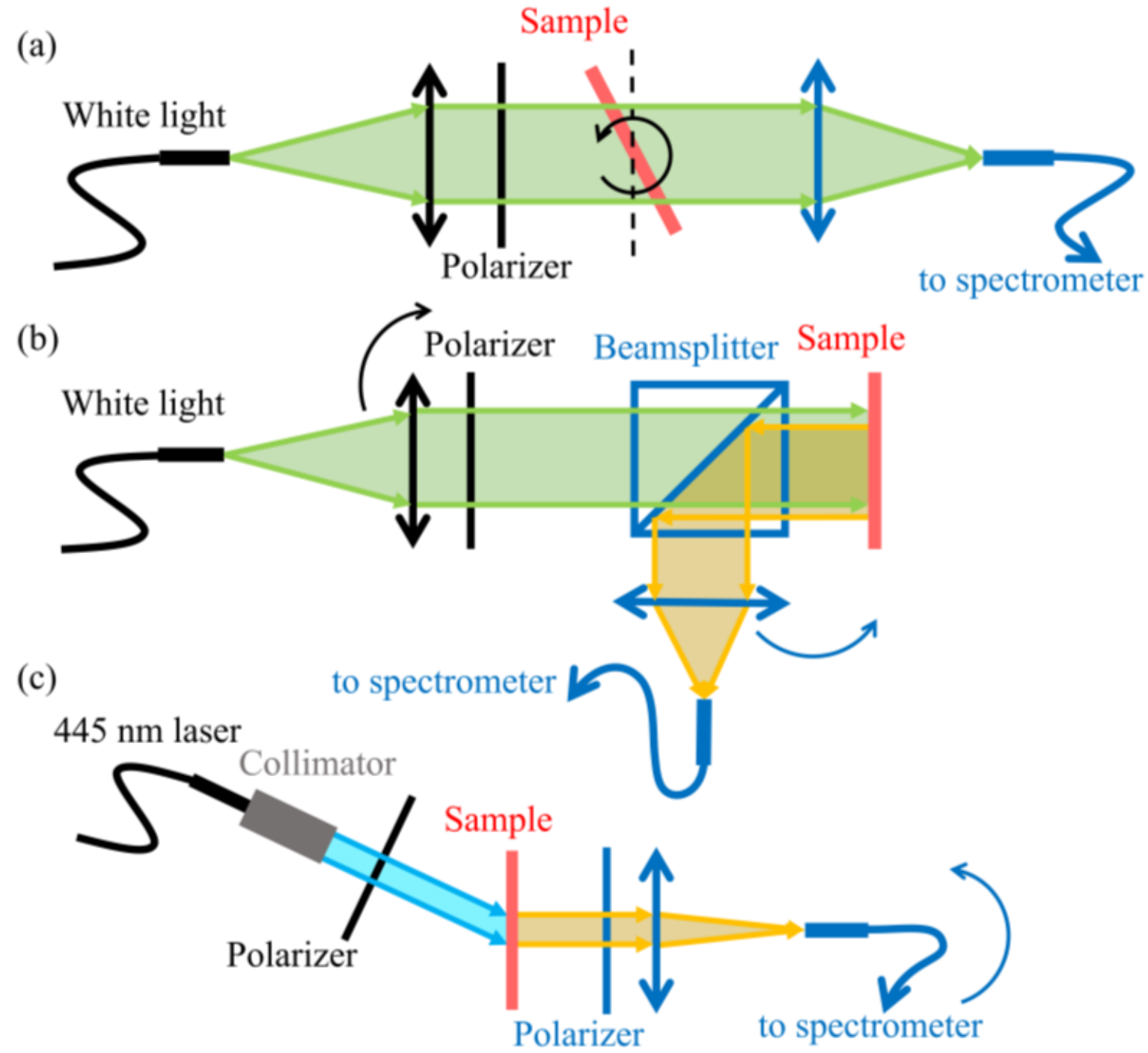

Figure S5. The optical measurement setups for (a) transmission, (b) reflection, and (c) photoluminescence measurements are illustrated. The optical elements of the incident side are drawn in black while that of the detection side are drawn in blue.

## S5. CALIBRATION OF THE ORGANIC DYES

The absorption and emission spectra of Lumogen F 305 red were measured as a reference. A PMMA layer (450 nm) with 1 wt.% of the Lumogen dye was spin-coated on an unstructured $SiO_2$ glass substrate. Fig. S6(a) shows the measured absorption and emission of the Lumogen dye normalized to the respective absorption and emission peak values.

We also measured the extinction coefficient $\kappa$ of the PMMA layer with IR780 iodide. A PMMA layer (450 nm) doped with 0.25 wt.% of IR780 iodide was spin-coated on an unstructured $SiO_2$ glass substrate. The normalized transmissivity of the dye layer was measured by UV/VIS/NIR spectroscopy (JASCO V-770) and was fitted by the Fresnel's Equations. The permittivity of the dye layer was modelled by the generalized Lorentz oscillator model with 6 oscillators.[49] As shown in Fig. S6(b), the best fit is consistent with the measured transmissivity, and the fitted $n$ and $\kappa$ of the dye layer are shown in Fig. S6(c).

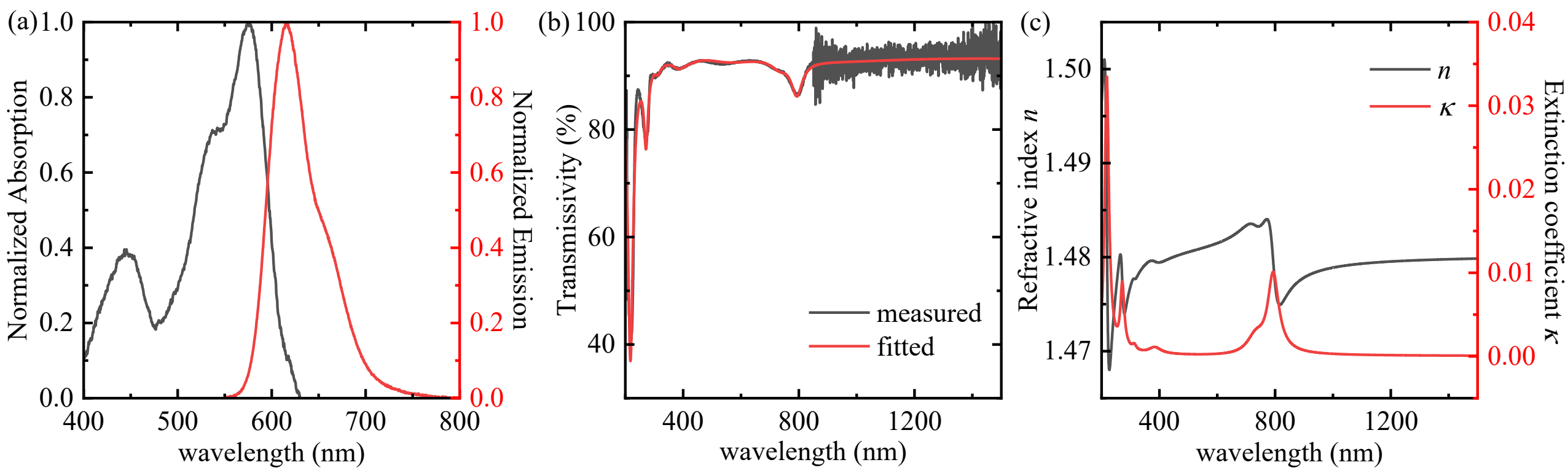

Figure S6. (a) The normalized absorption and emission spectra of Lumogen F 305 red. (b) The measured and fitted transmissivity of the PMMA layer (450 nm) with 0.25 wt.% of IR780 iodide is plotted against the wavelength. (c) The fitted $n$ and $\kappa$ of the PMMA layer with IR780 are plotted against the wavelength.